\documentclass{article}

\usepackage{arxiv}

\usepackage{hyperref}       
\usepackage{url}            
\usepackage{booktabs}       
\usepackage{amsfonts}       
\usepackage{nicefrac}       
\usepackage{microtype}      
\usepackage{lipsum}
\usepackage{graphicx}       
\usepackage{graphics}
\usepackage{cite}
\usepackage{subfig}

\usepackage{dblfloatfix}

\usepackage{enumitem}
\usepackage{amsmath}
\usepackage{amsfonts}
\usepackage{gensymb}
\usepackage{amssymb}
\usepackage{amsmath}
\usepackage{tikz}
\usepackage{pgfplots}
\usepackage{textcomp}
\usepackage{multirow}
\usepackage{siunitx}
\usepackage[mathscr]{euscript}
\usepackage{xspace}
\usepackage{enumerate}
\usepackage{hyperref}
\usepackage{url}
\usepackage{multirow}
\usepackage{tikz}
\usepackage{pgfplots}
\usepackage{textcomp}
\usepackage[utf8]{inputenc}
\usepackage{siunitx}

\title{A SMART RESOURCE MANAGEMENT MECHANISM WITH TRUST ACCESS CONTROL FOR CLOUD COMPUTING ENVIRONMENT 
}

\author{ Sakshi Chhabra$^a$, Ashutosh Kumar Singh$^b$ \\
$^a$Panipat Institute of Engineering and Technology, India \\
$^b$Department of Computer Applications, National Institute of Technology,\\
    Kurukshetra,  India\\
    $sakshichhabra555@gmail.com$, $ashutosh@nitkkr.ac.in$
}

\begin{document}
\maketitle

\begin{abstract}
The core of the computer business now offers subscription-based on-demand services with the help of cloud computing. We may now share resources among multiple users by using virtualization, which creates a virtual instance of a computer system running in an abstracted hardware layer. It provides infinite computing capabilities through its massive cloud datacenters, in contrast to early distributed computing models, and has been incredibly popular in recent years because to its continually growing infrastructure, user base, and hosted data volume. This article suggests a conceptual framework for a workload management paradigm in cloud settings that is both safe and performance-efficient. A resource management unit is used in this paradigm for energy and performing virtual machine allocation with efficiency, assuring the safe execution of users' applications, and protecting against data breaches brought on by unauthorised virtual machine access real-time. A secure virtual machine management unit controls the resource management unit and is created to produce data on unlawful access or intercommunication. Additionally, a workload analyzer unit works simultaneously to estimate resource consumption data to help the resource management unit be more effective during virtual machine allocation. The suggested model functions differently to effectively serve the same objective, including data encryption and decryption prior to transfer, usage of trust access mechanism to prevent unauthorised access to virtual machines, which creates extra computational cost overhead.
\end{abstract}

\keywords{Cloud Computing \and Resource utilization \and Trust access control \and Security \and Load balancing \and Resource allocation \and VM placement}

\section{Introduction}
The demand for high-end computing equipment has steadily increased over the last few years, which has caused the computational world to rapidly change \cite{1,2,3,4,5,J1}. New computational paradigms including cloud computing, grid computing, and cluster computing have emerged as a result of this progression. Cloud computing has become one of these concepts that has experienced substantial growth \cite{6,7,8,9,10,11,12,13,14,J2,15,16}. Three types of cloud computing are available: public cloud, private cloud, and hybrid cloud.  Additionally, a variety of services are offered that fall essentially into three categories: platform as a service (PaaS), infrastructure as a service (IaaS), and software as a service (SaaS) The foundation of cloud computing is Service-Oriented Architecture (SOA), which employs distributed computing and virtualization technologies \cite{17,18,19,20,21,22,23,24,25,26,27,28,29}. In SOA for cloud computing, users can access a shared pool of resources across a network and customise the available resources to meet their needs. Resource Management is a crucial component of virtualization and cloud computing (RM). Resource management (RM) is a procedure that handles the acquisition and release of resources. Flexible and on-demand resource provisioning is achieved via virtualization methods \cite{30,31,32,33}. To do this, either a new VM is built for each job that is received or it is added to the user's already-existing VM. When the job is finished, all the resources are freed and added to the pool of available resources. According to the Service Level Agreement (SLA) that is established between the service provider and client, resource assignment is carried out.
\begin{figure} [htbp]
	\centering
	\includegraphics[clip,trim=0cm 0cm 0cm 0cm, width=0.75\linewidth, height=6.5cm]{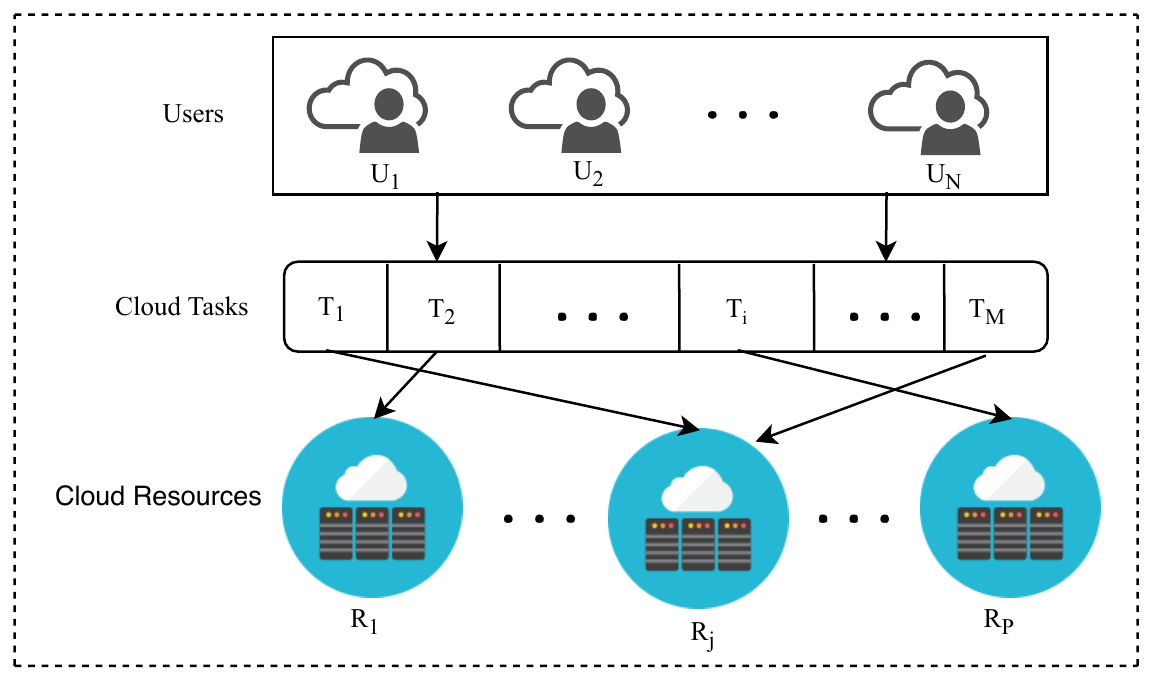}
	
	\caption{Cloud Resource scheduling View}
	\label{Load Balancing Categorization1}
\end{figure}
Resource management and consolidation are two key components of virtualization technologies used in cloud. More isolated physical computers may be combined into a virtualized environment using hypervisors in a cluster setting, using fewer physical resources than ever before \cite{34,35,36,37,38,39,40,41,42,43,44,45,46,47,48,49,50}.
Even while the situation is improved, this frequently falls short. Megawatts of electricity and thousands of physical computers are needed for large-scale cloud installations \cite{51,52,53}. Therefore, it is necessary to develop an effective Cloud computing system that makes the most of the Cloud's advantages while reducing its energy footprint. However, by utilising cloud services to share the resources, this can facilitate the system open to snooping, spoofing, and denial-of-service assaults (DoS)
assaults, trust issues etc. \cite{54,55,56,57,58,59,60}. Because of this, the number of flaws in cloud computing is increasing between the cloud service providers and the end clients.\\

\textbf{Why we adopt efficient Resource Management?}
	Resource management enables managers to make the most of the personnel, resources, and equipment used on their projects. Here are four compelling arguments for the process' importance in successful project management \cite{61,62,63,64,65,66,67,68,69}. It's significant because:
\begin{itemize}
        \item   Enhance resource utilisation: You may learn important information about how your resources will be used through resource management analysis and reporting.
    \item  Measures the effectiveness of the project: Making decisions based on resource management makes it clear what is required to carry out new initiatives. You can plan more effectively and increase project efficiency by tracking resource usage and ROI.
    \item 	Improve skill distribution: Resource management enables managers to predict resource demand more precisely. Resource managers can find areas for capacity and demand improvement by examining the current skills capacity and comprehending the future demand.
    \item Reduce the unnecessary cost: Resource management brings order to the chaos by providing managers with the tools they need to maximise their resources and reduce unnecessary spending. The data needed to analyse project and staff costs and enhance planning and budgeting is already available. A perfect team may also be maintained by distributing resources across tasks in accordance with capabilities and skill sets \cite{70,71,72,73}.
\end{itemize} 
The two core stages of resource management in cloud computing are resource provisioning and resource scheduling. Cloud resource provisioning is the process of making virtualized resources available for user allocation. Fig. 2 illustrates the load balancing is categorized as:
\begin{figure} [htbp]
	\centering
	\includegraphics[clip,trim=0cm 0cm 0cm 0cm, width=0.75\linewidth, height=6.5cm]{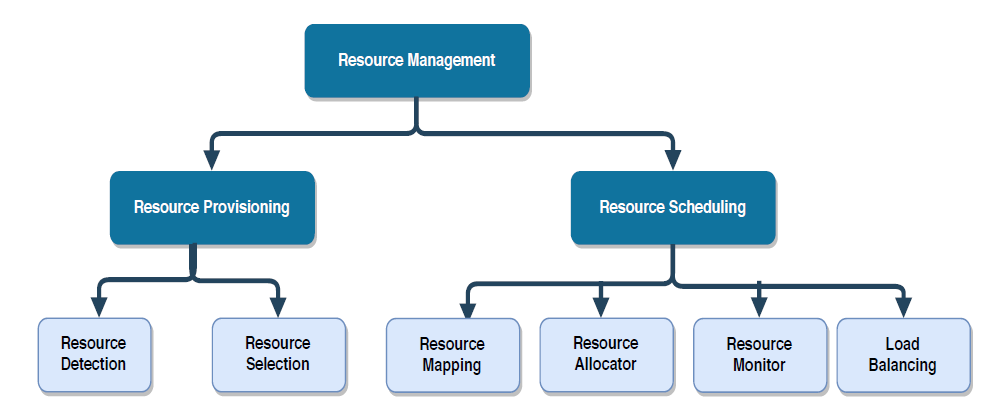}
	
	\caption{Resource Management Categorization}
	\label{Load Balancing Categorization1}
\end{figure}
A cloud service provider creates the appropriate number of virtual machines and distributes them among customers in line with their demands when a user requests resources from it. In compliance with quality of service (QoS) criteria, SLA agreements, and resource matching to projected workloads, resource provisioning is also responsible for providing for user demands. Scheduling may be used to choose the appropriate activity based on the necessary quality of service (QoS) requirements.\\
The storage of data and the hosting of numerous users present significant security vulnerabilities in the context of the cloud. The user data is now protected in the cloud by powerful technologies \cite{74,75,76}. In the cloud computing environment, this is becoming more complicated because to the increased security threats associated with the traffic transmitted through nodes. For instance, a hacker might introduce malicious software, which in turn could exploit a flaw and harm or lower the network's quality of service \cite{77,78,79,80,81,82,83,84,85,86,87,88,89,90}. In reality, one hacked cloud user can operate as a possible entry point for a man-in-the-middle attack (MITM), disrupt all connected users, leak data, utilise the service excessively, and harm the client's confidential data \cite{91,92,93,94,95,96,97,98.99,100,101,102,103,104,105,106,107}.
Therefore, the main problem of cloud-based IoT network is the design and construction of a powerful system that can effectively guarantee security and privacy without compromising and losing performance. Since lives are at risk, defence against these attacks is crucial. Controlling user access and keeping an eye on the entire system may be the most efficient way to solve such issues. To enable high-performance computing, the security and privacy procedures need to be examined with efficient resource utilization. \\
A smart and efficient resource management system is proposed, supported by a set of defined security rules and a set of resource management strategies, is required to solve the aforementioned issues in order to increase the system's intended security and performance. Collaborative nodes are used to monitor the many traffic kinds in cloud computing that are integrated to support IoT devices. The model will control a predetermined number of
cloud nodes To safeguard the network from attackers when a user is logged in, the proposed component comprises a monitoring agent that assesses and updates the trust level of connected users. Additionally, a certificate is sent to this user as a trust authorisation if the trust level is upheld during user access. In addition, it is suggested that resources be managed in an efficient
find appropriate resources based on SLA, while the computing resource
nodes have finite capacity constraints.\\
The remainder of the manuscript is structured as follows: A smart resource-aware management model and other relevant studies in the literature is presented in Section 2. The whole procedure of trust based access control mechanism is explained based on SLA and real-time resource status in Section 3 of this article. In Section 4, we go through the highlights of efficiency of our model, conclusion as well as future scope. 
\section{A Smart Resource-Aware Allocation Model}
The architecture of configuring, resource-aware, optimising, allocating and safeguarding policies for effective resource management is covered in this section. The layout of our proposed model can be shown in the Fig. 3.
The SLA-formatted descriptions with optimized resource allocation based on an intelligent resource-aware information manager. This conceptual model
is the primary safeguard that guarantees cloud service providers
can fulfil a lot of demands without going against SLA
based on QoS criteria and dynamically controls the resources
features of the user and the workload mentioned needs. Through the cloud workload management portal (CWMP), cloud consumers attempt to run the workloads. 
 The procedure of authenticating and authorising cloud consumers is then completed. 
\begin{figure} [htbp]
	\centering
	\includegraphics[clip,trim=0cm 0cm 0cm 0cm, width=1\linewidth, height=10.5cm]{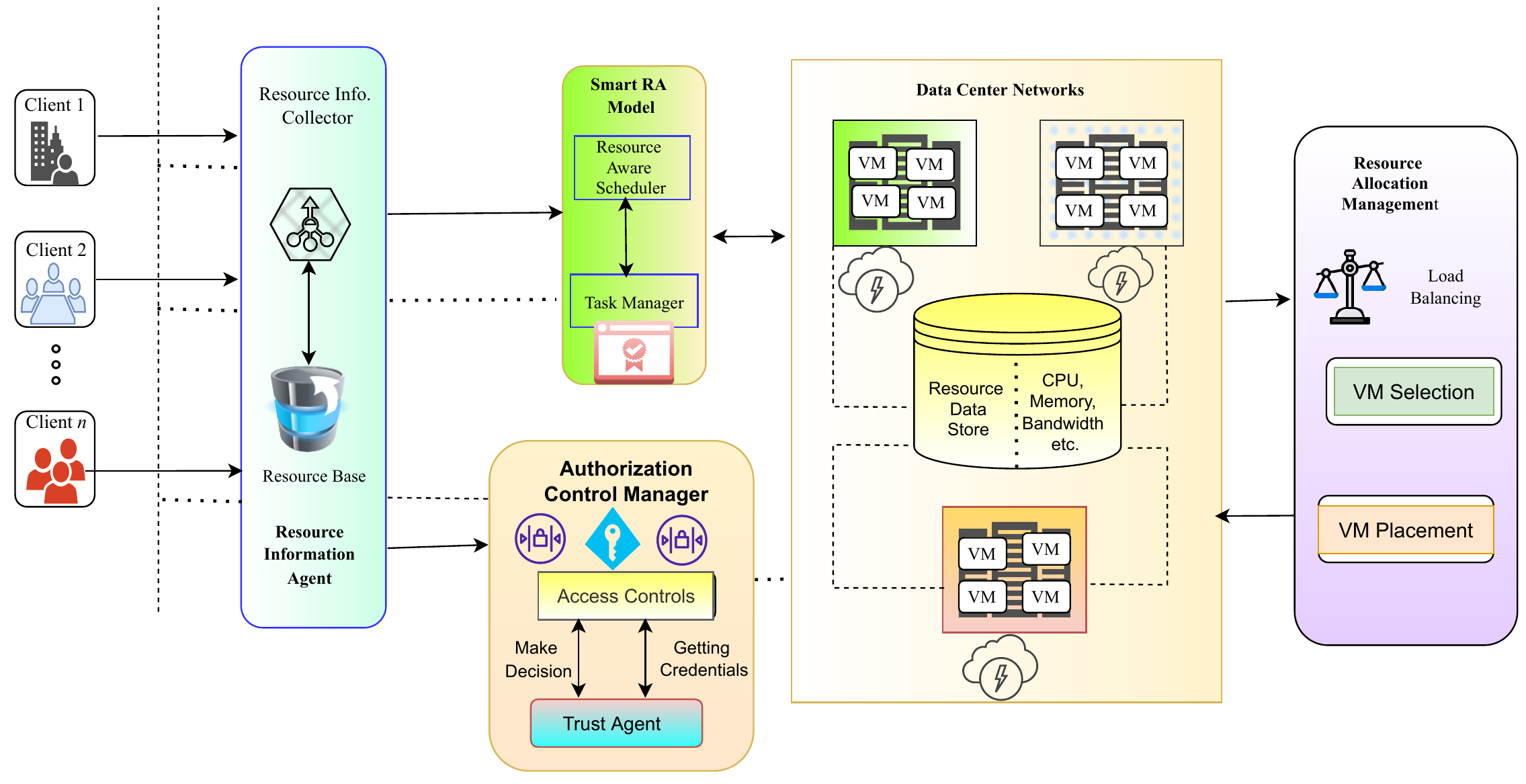}
	
	\caption{Proposed Architecture}
	\label{Load Balancing Categorization1}
\end{figure}

 Following authentication, this model requests that the cloud consumer needs (SLA) be submitted, which the authorised cloud consumer does. They next submit a request for the availability of a specific resource with the necessary details for the execution of their task. It analyses the numerous workload details that the cloud consumer requested before taking the information from the relevant task. For multi-tenancy, we took into account various cloud providers that communicate with one another and update their new guidelines and regulations for cloud resources.\\
 The purpose of Workload Manager is to examine several aspects of a cloud workload to ascertain whether or not it is feasible to move an application to the cloud. The various cloud workloads each have their own set of specifications and characteristics. This study offers guidance for the execution strategy as well.
It is divided into three sub-units: the workload queue, the workload description, and the bulk of the tasks. Every task that a cloud consumer submits to be executed is taken into account.
 In the first stage, a client requests that a job be executed on resources that can satisfy the client's QoS requirements. The resource information collector receives the assignment from the clients to determine if any resources satisfy the specified QoS restrictions and provide a profit. The QoS parameters are used by the Smart RA model service to identify the best, operating resource.

 Both the runtime estimator and the profit estimator get in touch with the RIC and ask for information about each resource's state in order to calculate runtime and profit. Both estimators receive information about the state of the resources once the RIC searches its resource information base. The admission controller examines the estimated runtime and profit values after runtime and profit have been calculated in order to identify resources that are suitable for meeting SLA requirements. Only for divisible jobs, the divisible job scheduler makes a policy to reduce overall runtime through the job division in the event that adequate resources cannot be found.
whether the newly chosen resources meet the user's needs or not. The architectural view of proposed model is illustrated in Fig. 3.\\
Numerous users from all around the world submit requests for programme execution to a resource management unit. These applications may be submitted in any format imaginable, including those for high-performance computing, scientific processes, the military, the pharmaceutical or research industry, internet commerce and marketing, etc. These programmes are split up by the resource management unit into one or more tasks or components that can run simultaneously on several virtual machines. A virtual network of virtual machines with permitted access connections is created when all the virtual machines executing a component of a shared application need to communicate and exchange information with one another. This information about permitted virtual machine links is kept and updated by the resource management unit in a resource information agent.\\
The current method demonstrates a generic cloud computing environment and the function of resource management inside a cloud environment, where cloud architecture consists of groups of servers connected by router networking devices to exchange information as needed. Numerous physical servers with the same or different resource capacities make up each cluster. The resources might consist of a CPU, RAM, and other things.
RAM on the disc, bandwidth, etc. For the purpose of handling requests from cloud user applications, virtual machines—which might take the shape of compute, storage, or networking instances—are hosted on servers. The resource management unit receives application requests from the cloud users $$(User 1, User 2,..., User n)$$ for processing at the cloud infrastructure. Applications containing requests are sent to the Resource Management Unit by different cloud users $$(User 1, User 2,..., User n)$$\\
Additionally, it simplifies the management of virtual machines, including scheduling because of the ineffective resource distribution across virtual machines, placement, migration, and management of over- or under-loading scenarios. Additionally, it assists with task management by carrying out the work of breaking down user applications into smaller components or tasks, assigning these tasks to specific virtual machines, and scheduling their execution. The response for the user retrieved via Resource Management Unit is produced by integrating the output processed tasks of an application. The new/unknown as well as known attacker virtual machines allow the suggested methodology to prevent data security threats. Different virtual machines that are installed on varied servers around the datacenter are given different duties by the resource management unit. The CPU, memory, disc, and bandwidth consumption during task execution by various virtual machines are tracked, and a workload/resource use data store is established and updated on a regular basis. A machine learning-based workload analyzer is trained and retrained using the data stored in this data store as input.
This workload analyzer is used to forecast future resource usage on various servers and virtual machines. In order to effectively manage load, the resource management unit uses estimated resource usage information to make a variety of proactive decisions pertaining to load management, such as creating, terminating, or migrating different virtual machines, handling over/under-load conditions of servers, etc.

\section{Trust-based Access Control Mechanism}
The storage of data and the hosting of numerous users present significant security vulnerabilities in the context of the cloud. The user data is now protected in the cloud by powerful tools \cite{108,109,110,111,112,113,114,115,116,117,118,119,120}. The cloud computing environment is becoming more complex due to the increased security threats associated with traffic transmitted over nodes. 

\begin{figure} [htbp]
	\centering
	\includegraphics[clip,trim=0cm 0cm 0cm 0cm, width=1\linewidth, height=11cm]{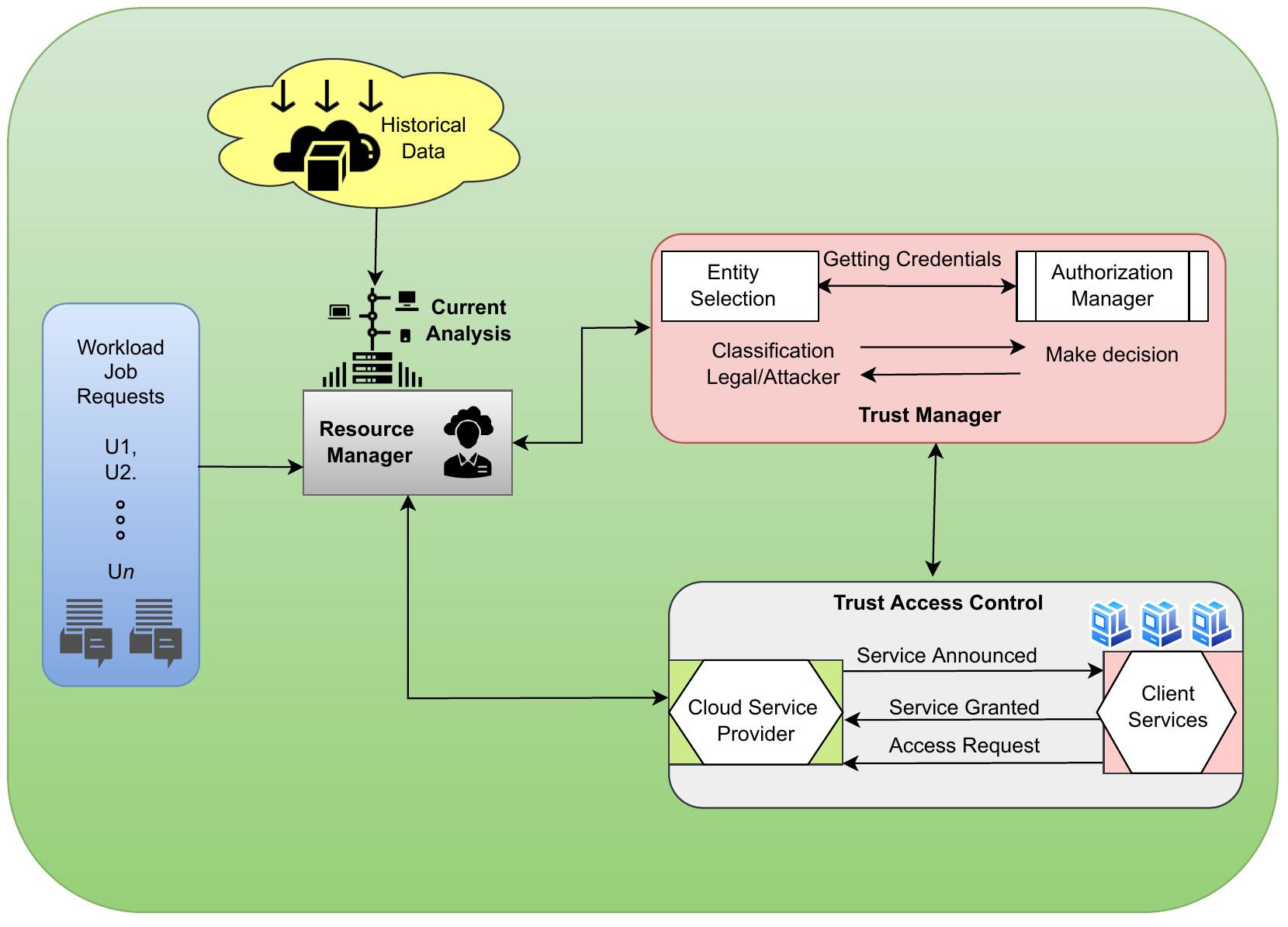}
	
	\caption{Trust-based Secure Mechanism}
	\label{Load Balancing Categorization1}
\end{figure}

For illustration, a hacker may use malevolent apps may take advantage of a vulnerability to harm or
decrease network quality of service \cite{121.122,123}. In actuality, a single hacked node
can create a Man-in-the-Middle attack (MITM) access point, disrupt all connected users, leak data, abuse the service, and corrupt data in the cloud nodes. A rogue internal user might launch an MITM attack, which could endanger the cloud network by intercepting, hacking, injecting, and filtering incoming data from end users. Therefore, the main problem of a wireless cloud-based IoT network is to design and build a powerful system that can effectively guarantee security and privacy without sacrificing and losing speed. Since lives are at risk, defence against these attacks is crucial. Controlling user access and keeping an eye on the entire system may be the most efficient way to solve such issues. To enable high performance computing, the security and privacy procedures need to be examined. Due to their limited resources, cloud nodes find it extremely difficult to handle numerous queries at once. In this situation, the computing performance of the cloud can be significantly reduced. In addition, providers might not be able to reach the desired performance without effectively securing access to cloud node resources in an IoT network.\\
The security system's implementation can go wrong and cause major problems. In light of this, it is crucial to choose the security methods to be used, the necessary modules/elements, and the established performance standards carefully.
To the best of our knowledge, improving a system's security does not always entail degrading its speed. For instance, security methods should be established as a fundamental component of the cloud ecosystem because if they weren't, their performance would be affected by assaults like malware, resource abuse, etc. IoT gadgets and wireless sensors make up the bulk of cloud systems. The wireless network gives attackers unprecedented flexibility to interrupt and intercept sensitive data being shared if it is not disguised and encrypted. However, we provide resource management strategies and security measures for cloud computing in order to decrease the system's overall management costs by taking into consideration diverse and flexible activities.\\
According to the suggested model's embodiment, Fig. 4 depicts a full perspective of a trust-based secure mechanism and performance-efficient load management model in a cloud environment. By submitting requests to the resource management unit in the form of various apps, cloud customers are able to take use of cloud infrastructure services. According to the task management, these applications are separated into tasks, which are then allocated to execute on various virtual machines depending on whether the task's resource needs are less than or equal to those of the virtual machine, as managed by the virtual machine management.

The objective of this paper is to manage each new cloud user's access. It determines the trust level of such a system access request. The cluster head then assigns each user a specific authorisation that includes information about his trust level.
The resource allocation process is followed by the usage of the authorisation. This user doesn't restart the access process on his or her subsequent access in that little cell since they already have authorization, they merely need to present it.
\section{Conclusion and Future scope}
In order to answer the concerns that customers have regarding quality, efficiency, and maximising the fulfilment of their demands, these new cloud-based solutions first determine the security and privacy standards that are now in effect. We thoroughly grasp the significance of secure cloud computing and resource effectiveness in order to evaluate our ideas. Load balancing helps to improve workload allocation across various computer resources because security helps to avoid data leaks and disposals. The planning that is being considered or suggestion relates to:

\begin{itemize}
    \item [$\checkmark$] For increased scalability in cloud computing, machine learning algorithms may be utilised to forecast the incoming data rate or application demands.
    \item [$\checkmark$] Meta-heuristic optimization can be used to augment some of the models, which are heuristic optimization models.
    \item [$\checkmark$] The suggested technique may be utilised to enhance various QoS criteria for better cloud services, such as elasticity, predictability, security, performance, SLA violation, and response time.
    \item [$\checkmark$] LB algorithms that have been developed can be evaluated in the future at actual processes as Montage, EpiGenomics, CyberShake, LIGO, and SIPHT.
    \item [$\checkmark$] Other learning paradigms, such as deep learning, learning influenced by quantum mechanics, capsule networks, etc., can be used to construct the load balancing models.
    \item [$\checkmark$] The degree of confidence in cryptographic services is increased by decision-makers choosing the best cloud service provisioning and using it in practise there. \cite{1}.
\end{itemize}

\bibliographystyle{IEEEtran}
\newpage \bibliography{main}

  

\end{document}